\documentclass[a4paper, 11pt]{article}

\title{The Vlasov equation cannot fully account for collisionless shocks}
\author{Antoine Bret\footnote{ETSI Industriales, Universidad de Castilla-La Mancha, 13071 Ciudad Real, Spain\\
Instituto de Investigaciones Energ\'{e}ticas y Aplicaciones Industriales, Campus Universitario de Ciudad Real,  13071 Ciudad Real, Spain}}
\date{}

\usepackage{graphicx}
\usepackage[T1]{fontenc}
\usepackage{numprint}
\usepackage{lmodern}
\usepackage[utf8]{inputenc}
\usepackage{color}

\begin{document}

\maketitle

\begin{center}
\textbf{Abstract}
\end{center}
  It is argued that the Vlasov equation cannot fully account for collisionless shocks since it conserves entropy, while a shock does not. A rigorous mathematical theory of collisionless shocks could require working at the Klimontovich level.

  \newpage

\section{Introduction}
Shock waves are fundamental processes in fluids and plasmas. In a fluid, the shock transition is mediated by binary collisions. In the absence of binary collisions, namely when the mean free path is much larger than the dimensions of the system, it was found from the 1960s that shock waves can be sustained by collective electromagnetic effects in a collisionless plasma \cite{Sagdeev66}. Such shocks have been dubbed ``collisionless shocks''.

Due to their omnipresence in astrophysics, collisionless shocks have been under intense scrutiny over the last decades, experimentally, numerically and theoretically (see \cite{Marcowith2016} and references therein). Their ability to accelerate particles with a power law spectrum makes them good candidates to explain the origin of the cosmic rays detected on earth \cite{Blandford78,Cronin1999}. Studying them in laboratory remains challenging as it takes resources  like the National Ignition Facility to produce such shocks an observe particles acceleration \cite{Fiuza2020}.

From the numerical point of view, Particle-In-Cell simulations (PIC) are well adapted to simulate them \cite{Spitkovsky2005,Pohl2020PrPNP}. Such simulations simulate the particles motions from basic principles:  a given, discrete, particles distribution is used to derive the charge and current distribution, which, in turn, are used to advance each particle \cite{birdsall2004}. The physics involved is therefore fundamental (only Maxwell's equations and Newton's laws), with the algorithm operating at the microscopic level.

Yet, the longest simulations to date can only probe a glimpse of the history of a real system \cite{Keshet09,Groselj2024}. Theoretical approaches are therefore still much needed.

\begin{figure}
\begin{center}
 \includegraphics[width=\textwidth]{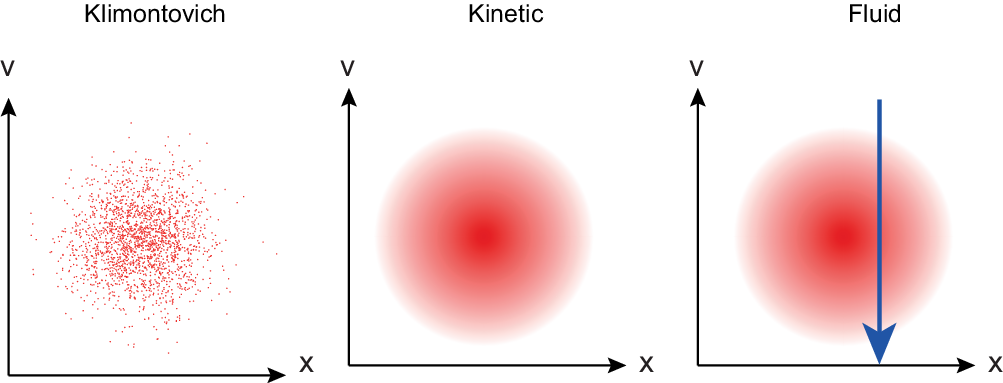}
\end{center}
\caption{Three levels of description of a 1D system of particles. Left, complete knowledge of each position and velocity. Center, continuum approximation of the latter. Right, fluid approach based on moments (represented by the blue arrow) in velocity space.}\label{fig}
\end{figure}

\section{The issue with the Vlasov equation}
In the absence of binary collisions, it is usually considered that collisionless shocks should be theoretically dealt with at the kinetic level, via de Vlasov equation.  Yet, this equation conserves entropy (\cite{LandauKinetic}, \S 27), while a shock, collisionless or not, does not \cite{Zeldovich}. Such entropy increase is simply the result of conservation of matter, momentum and energy through the shock front, which impose a downstream state with entropy higher than the upstream. See for example Ref. \cite{landaufluid}, \S 85.

Regarding astrophysical settings, entropy generation has been measured across the Earth's bow shock, and found consistent with a Vlasov model of the entropy \emph{density} (not the \emph{total} entropy, which Vlasov conserves) \cite{Parks2017RvMPP,ParksPRL2012}.

What is then the issue with the Vlasov equation?

Figure \ref{fig} gives a hint. Consider for simplicity $N$ particles in a 1D space\footnote{The reasoning can be readily adapted to a time dependent problem in 3D.}, with position $x_i$ and velocity $v_i$. Time dependence is not mentioned but can be included seamlessly. Three levels of description are possible.

\begin{itemize}
  \item The finest possible description of the system consists in knowing the velocity $v_i$ and position $x_i$ of each of the $N$ particles \cite{nicholson1983}. Let us call it the ``Klimontovich level'',  sketched on the left, with a distribution of the form,
\begin{equation}\label{eq:klim}
F(x,v)=\sum_i \delta(x-x_i)\delta(v-v_i).
\end{equation}
The number $N_{\Delta x,v}$ of particles in the phase space volume $\Delta x \Delta v$ is given by,
\begin{equation}\label{eq:Nklim}
N_{\Delta x,v} = \int_{\Delta x}\int_{\Delta v}F(x,v)dx dv,
\end{equation}
and will be 0 or 1 when $\Delta x,v \rightarrow 0$, depending on whether there is a particle where the 0 limit is considered, or not. At any rate, Eq. (\ref{eq:Nklim}) can only yield an integer.
  \item Then, sketched at the center, the ``kinetic level''\footnote{Note that ``kinetic'' usually refers in literature to both the present ``Klimontovich'' and ``kinetic'' levels.} consists in a continuum approximation of the discrete distribution pictured on the left. It assumes that the phase space $(x,v)$ is populated densely enough for a continuum description to be valid. There, $f(x,v)dxdv$ gives the number particles in the phase space element $dxdv$, where $f(x,v)$ is the distribution function.

      The integration of $f(x,v)$ over a phase space volume $\Delta x \Delta v$ gives a real number.
  \item Finally, right sketch, the fluid description proceeds through moments (represented by the blue arrow) in velocity space of the distribution function.
\end{itemize}

\section{Conclusion}
The Vlasov equation is only likely to break down for a collisionless shock due to the continuum hypothesis, necessary to go from the ``Klimontovich'' to the ``Kinetic'' level. At one point of the motion along the shock,  probably about the shock front, the Klimontovich distribution, that is, the real distribution, has to become so sparse that the continuum hypothesis fails. As a result, entropy can grow when Vlasov would not let it.

Noteworthily, PIC simulations do work at the Klimontovich, particle, level, which explains how they can simulate collisionless shock waves.
Operating at the particle level, using only Maxwell’s and Newton’s equations to advance the particles, they do not solve the Vlasov equation, being thus free from its continuum approximation in the $(x,v)$ phase space. Indeed, entropy increase in PIC simulations of shocks has been computed from first principles, from the simulation data, for example in Ref. \cite{Haggerty2022}.

It seems therefore that elaborating a rigorous mathematical theory of collisionless shocks requires working at the Klimontovich level as well.
An option could be to write the Klimontovich function (\ref{eq:klim}) as \cite{Schroedter2024},
\begin{equation}
F(x,v) = f(x,v) + \delta F(x,v),
\end{equation}
where $ f(x,v) \equiv <F(x,v)>$ is the kinetic distribution function, and $\delta F(x,v)$  the fluctuation around the mean. Inserting then this expression into the Vlasov equation yields a nonlinear term with the induced field $E$ of the type $< \delta F \delta E>$, representing particles-particles collisions, plus particles scattering by the field (see Ref. \cite{Schroedter2024} for a recent discussion).
Alternatively, a formalism such as the one presented in Ref. \cite{Ewart2024} could be explored in relation to the present problem.

\section{Acknowledgments}
The author acknowledges support from the Spanish Ministerio de Econom\'{\i}a y Competitividad  (Grant No. PID2021-125550OB-I00). Thanks are due to John Kirk for enriching discussions.

 \bibliographystyle{abbrv}
\bibliography{BibBret}

\end{document}